\documentclass[a4paper,11pt]{article}

\usepackage{pos}

\usepackage{amsmath}

\usepackage{graphicx}
\usepackage{slashed}
\usepackage{booktabs,tabulary}
\usepackage{placeins}

\newcommand{\mpii}{M_{\pi^0}}
\newcommand{\mpic}{M_{\pi^\pm}}
\newcommand{\beq}{\begin{equation}}
\newcommand{\eeq}{\end{equation}}
\newcommand{\Order}{\mathcal{O}}
\newcommand{\GeV}{\,\text{GeV}}
\newcommand{\MeV}{\,\text{MeV}}

\renewenvironment{thebibliography}[1]{
  \begin{oldthebibliography}{#1}
    \setlength{\itemsep}{0.5em}
    \setlength{\parskip}{0em}
}
{
  \end{oldthebibliography}
}

\allowdisplaybreaks
\begin{document}

\title{Chiral extrapolation of hadronic vacuum polarization and isospin-breaking corrections}

\ShortTitle{Chiral extrapolation of HVP and isospin-breaking corrections}

\author*[a]{Martin Hoferichter}

\author[a]{Gilberto Colangelo}

\author[a]{Bai-Long Hoid}

\author[b]{Bastian Kubis}

\author[c]{\\Jacobo Ruiz de Elvira}

\author[b]{Dominik Stamen}

\author[d,e]{Peter Stoffer}

\affiliation[a]{Albert Einstein Center for Fundamental Physics, Institute for Theoretical Physics, University of Bern, Sidlerstrasse 5, 3012 Bern, Switzerland}

\affiliation[b]{Helmholtz-Institut f\"ur Strahlen- und Kernphysik (Theorie) and
Bethe Center for Theoretical Physics, Universit\"at Bonn, 53115 Bonn, Germany}

\affiliation[c]{Universidad Complutense de Madrid, Facultad de Ciencias F\'isicas,
Departamento de F\'isica Te\'orica and IPARCOS, Plaza de las Ciencias 1, 28040 Madrid, Spain}

\affiliation[d]{Physik-Institut, Universit\"at Z\"urich, Winterthurerstrasse 190, 8057 Z\"urich, Switzerland}

\affiliation[e]{Paul Scherrer Institut, 5232 Villigen PSI, Switzerland}

\emailAdd{hoferichter@itp.unibe.ch}
      
\abstract{By far the biggest contribution to hadronic vacuum polarization (HVP) arises from the two-pion channel. Its quark-mass dependence can be evaluated by combining dispersion relations with chiral perturbation theory, providing guidance on the functional form of chiral extrapolations, or even interpolations around the physical point. In addition, the approach allows one to estimate in a controlled way the isospin-breaking (IB) corrections that arise from the pion mass difference. As an application, we present an updated estimate of phenomenological expectations for electromagnetic and strong IB corrections to the HVP contribution to the anomalous magnetic moment of the muon. In particular, we include IB effects in the $\bar K K$ channel, which are enhanced due to the proximity of the $\bar K K$ threshold and the $\phi$ resonance. The resulting estimates make it unlikely that the current tension between lattice-QCD and data-driven evaluations of the HVP contribution is caused by IB corrections. 
 }

\FullConference{The 39th International Symposium on Lattice Field Theory,\\
8th-13th August, 2022,\\
Rheinische Friedrich-Wilhelms-Universität Bonn, Bonn, Germany}

		\maketitle
		
\section{Introduction}

Understanding the tension between data-driven~\cite{Aoyama:2020ynm,Davier:2017zfy,Keshavarzi:2018mgv,Colangelo:2018mtw,Hoferichter:2019gzf,Davier:2019can,Keshavarzi:2019abf,Colangelo:2022vok} and lattice-QCD determinations~\cite{Borsanyi:2020mff,Ce:2022kxy,Alexandrou:2022amy} of the hadronic-vacuum-polarization (HVP) contribution to the anomalous magnetic moment of the muon is of critical importance for the interpretation of the current $4.2\sigma$
discrepancy between experiment~\cite{Muong-2:2006rrc,Muong-2:2021ojo,Muong-2:2021ovs,Muong-2:2021xzz,Muong-2:2021vma} and the prediction in the Standard Model~\cite{Aoyama:2020ynm,Aoyama:2012wk,Aoyama:2019ryr,Czarnecki:2002nt,Gnendiger:2013pva,Davier:2017zfy,Keshavarzi:2018mgv,Colangelo:2018mtw,Hoferichter:2019gzf,Davier:2019can,Keshavarzi:2019abf,Hoid:2020xjs,Kurz:2014wya,Melnikov:2003xd,Masjuan:2017tvw,Colangelo:2017qdm,Colangelo:2017fiz,Hoferichter:2018dmo,Hoferichter:2018kwz,Gerardin:2019vio,Bijnens:2019ghy,Colangelo:2019lpu,Colangelo:2019uex,Blum:2019ugy,Colangelo:2014qya}, using $e^+e^-\to\text{hadrons}$ cross-section data for the latter. The comparison is well-defined for the total HVP contribution as well as for windows in Euclidean time~\cite{Blum:2018mom}, but, to some extent, even partial quantities evaluated in lattice QCD can be subject to further independent cross checks. In these proceedings, we focus on the role of isospin-breaking (IB) corrections. These have been estimated from phenomenology before~\cite{Borsanyi:2017zdw,Jegerlehner:2017gek}, but with recent work on the dominant exclusive channels using dispersion relations and chiral perturbation theory (ChPT)~\cite{Colangelo:2018mtw,Hoid:2020xjs,Colangelo:2020lcg,Colangelo:2021moe,Colangelo:2022prz} several estimates can be improved, most notably the impact of the pion-mass difference on the $2\pi$ channel. In addition, we include the $\bar K K$ channel~\cite{Stamen:2022uqh}, in which case IB effects in the kaon mass are enhanced due to the proximity of the $\bar K K$ threshold and the $\phi$ resonance. All numbers will be given in units of $10^{-10}$.

\section{Pion-mass dependence of the two-pion channel}

\begin{figure}[t]
\centering
\includegraphics[width=0.49\linewidth]{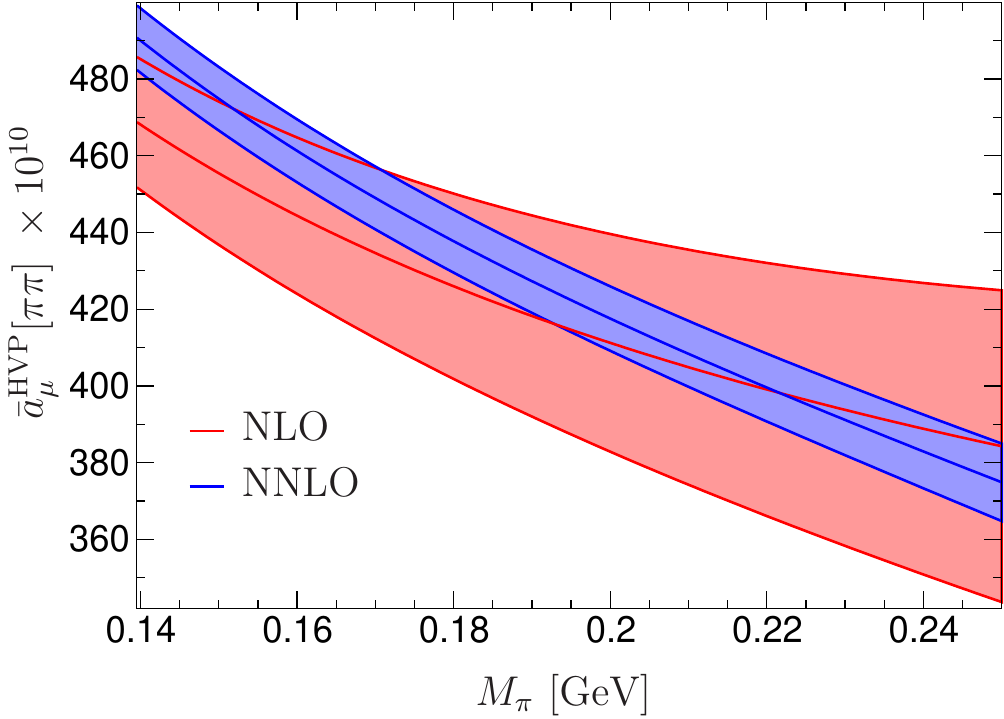}
\includegraphics[width=0.49\linewidth]{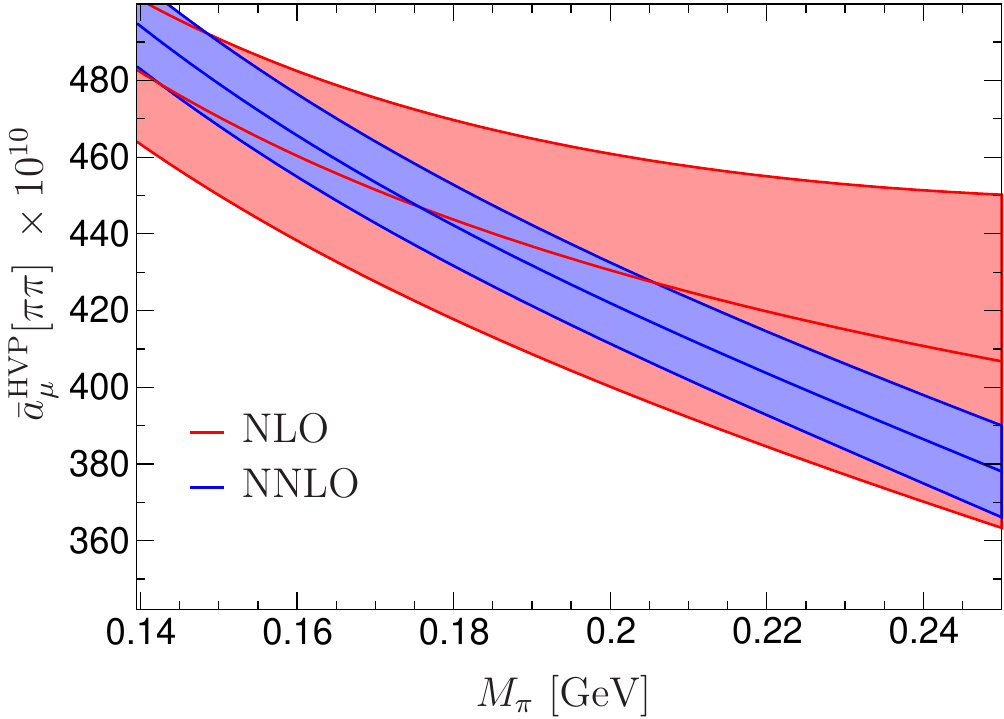}
\caption{Pion-mass dependence of $\bar a_\mu^\text{HVP}[\pi\pi]$ from the NLO (red) and NNLO (blue) IAM, for a normal (left) and conformal (right) polynomial. Figures taken from Ref.~\cite{Colangelo:2021moe}.}
\label{fig:chiral_extr}
\end{figure}

Throughout, we use a decomposition of the pion form factor
\beq
F_\pi^V(s) = \Omega_1^1(s)\times G_\omega(s) \times G_\text{in}(s),
\eeq
where the three factors incorporate two-pion, three-pion, and higher intermediate states, respectively. The Omn\`es factor $\Omega_1^1(s)$~\cite{Omnes:1958hv} does so in terms of the $P$-wave $\pi\pi$ scattering phase shift, $G_\omega(s)$ parameterizes $\rho$--$\omega$ mixing in terms of the residue $\epsilon_\omega$ at the $\omega$ pole, and $G_\text{in}(s)$ is expanded in a (conformal) polynomial, whose parameters can be matched onto the pion charge radius $\langle r_\pi^2\rangle$ and higher orders in the low-energy expansion of $F_\pi^V(s)$.  Given that $G_\omega$ already represents an IB effect, it suffices to study the pion-mass dependence of the pure $I=1$ correlator, denoted in Fig.~\ref{fig:chiral_extr} by $\bar a_\mu^\text{HVP}[\pi\pi]$ to indicate that $\epsilon_\omega=0$. To obtain the pion-mass dependence of the $\pi\pi$ phase shift and thus $\Omega_1^1(s)$~\cite{Guo:2008nc}, we employ the inverse amplitude method (IAM) at one- (NLO) and two-loop (NNLO) order~\cite{Niehus:2020gmf}, with parameters determined from a combined fit to lattice QCD~\cite{Andersen:2018mau} and phenomenology~\cite{Colangelo:2018mtw}. For $G_\text{in}(s)$, we use the known two-loop expansion of $\langle r_\pi^2\rangle$~\cite{Bijnens:1998fm}. Here, the main uncertainty arises from a new 
low-energy constant $r_{V1}^r=2.0\times 10^{-5}$, which we estimate from resonance saturation and validate with lattice-QCD calculations of $\langle r_\pi^2\rangle$ at larger-than-physical pion masses~\cite{Feng:2019geu,Wang:2020nbf}. The resulting  prediction for the pion-mass dependence in Fig.~\ref{fig:chiral_extr} reproduces the value at the physical point within uncertainties. Possible applications to lattice QCD are discussed in Ref.~\cite{Colangelo:2021moe}, ranging from a full fit of the $I=1$ contribution to tests of the strength of infrared singularities in the relevant fit region~\cite{Golterman:2017njs}. In the application to IB, we find that the difference between charged and neutral pion mass gives
\beq
a_\mu^\text{HVP}[\pi\pi]\big|_{\mpic}-a_\mu^\text{HVP}[\pi\pi]\big|_{\mpii}=-7.67(4)_\text{ChPT}(3)_\text{polynomial}(4)_{\langle r_\pi^2\rangle}(21)_{r_{V1}^r}[22]_\text{total},
\eeq
where the uncertainties refer to chiral convergence, comparison of a normal and conformal polynomial, and the uncertainties in $\langle r_\pi^2\rangle$, $r_{V1}^r$, respectively. This effect arises predominantly from the threshold region, in such a way that the resulting contribution is almost exclusively contained in the long-distance (LD) window. 
  
\section{$\boldsymbol{\rho}$--$\boldsymbol{\omega}$ mixing and final-state radiation}

Final-state radiation (FSR) is dominated by infrared enhanced effects, see Refs.~\cite{Ignatov:2022iou,Colangelo:2022lzg} for an explicit test in the context of the forward--backward asymmetry, and we will  adopt the results from Ref.~\cite{Colangelo:2022prz} obtained under this assumption, supplemented by small dispersive corrections from Ref.~\cite{Moussallam:2013una}
\beq
a_\mu^\text{HVP}[\pi\pi\gamma, \text{non-Born}]=0.15_{\pi^+\pi^-\gamma}+0.03_{\pi^0\pi^0\gamma}=0.18(4).
\eeq
With mixed higher-order terms  $\Order(e^2\epsilon_\omega)$ small, $\lesssim 0.1$, we have~\cite{Colangelo:2022prz}
\beq
a_\mu^\text{HVP}[\pi\pi, \text{FSR}, \text{Born}]=4.24(2),\qquad 
a_\mu^\text{HVP}[\pi\pi, \rho\text{--}\omega]=3.68(17).
\eeq
While the FSR contribution represents a pure $\Order(e^2)$ effect, it is less clear how to separate $\rho$--$\omega$ mixing into 
its QED, $\Order(e^2)$, and strong, $\Order(\delta)=\Order(m_u-m_d)$, parts. After removing a $\gamma$--$\omega$-mixing diagram that is subtracted in the bare cross section, 
leading-order vector-meson ChPT~\cite{Urech:1995ry} suggests that the entire effect should be of $\Order(\delta)$, but later work showed that higher-order corrections are difficult to estimate~\cite{Bijnens:1996nq,Bijnens:1997ni}. We will continue to book $\rho$--$\omega$ mixing in the $\Order(\delta)$ category, emphasizing that this ambiguity could potentially shift contributions between the two classes of IB.

\section{Isospin breaking in the $\boldsymbol{\bar K K}$ channel}

The threshold region in the $\bar K K$ channel is dominated by the isoscalar form factor, and can thus be analyzed in terms of the $\phi$ resonance parameters~\cite{Stamen:2022uqh}. The relevant IB effects arise from FSR, from IB in the kaon masses, and from IB in the $\phi$ residues
\beq
c_\phi^{K^+K^-}=0.977(6),\qquad c_\phi^{\bar K^0K^0}=1.001(6). 
\eeq
The latter gives the dominant contribution to the uncertainty, about $\simeq 0.8$ in the full HVP integral, as it is not clear which residue, or combination of the two, should be identified with the isospin limit. To define the kaon masses in the isospin limit, we use the charged-kaon self energy $(M_{K^\pm}^2)_\text{EM} = 2.12(18)\times 10^{-3}\GeV^2$ from the Cottingham formula~\cite{Stamen:2022uqh}, leading to
\beq
\label{mass_decomposition}
M_{K^\pm} = (494.58-3.05_\delta + 2.14_{e^2}\big)\MeV, \qquad 
M_{K^0} = (494.58+3.03_\delta\big)\MeV,
\eeq
which is close to typical quark-mass-scheme decompositions in lattice-QCD~\cite{Lellouch,Portelli}. Using the $\phi$ spectral function from Ref.~\cite{Stamen:2022uqh} and varying the kaon masses according to Eq.~\eqref{mass_decomposition}, we obtain 
\begin{align}
\label{KKbar}
%a_\mu^\text{HVP}[K^+K^-, \leq 1.05\GeV]&=18.45(20), &
%a_\mu^\text{HVP}[K^0\bar K^0, \leq 1.05\GeV]&=11.83(15),\notag\\
a_\mu^\text{HVP}[K^+K^-, \text{FSR}]&=0.75(4), & &\notag\\
a_\mu^\text{HVP}[K^+K^-, e^2]&=-3.24(17),& 
a_\mu^\text{HVP}[K^0\bar K^0, e^2]&=-0.02(0),\notag\\
a_\mu^\text{HVP}[K^+K^-, \delta]&=4.98(26), &
a_\mu^\text{HVP}[K^0\bar K^0, \delta]&=-4.62(23),\notag\\
a_\mu^\text{HVP}[K^+K^-, e^2\delta]&=-0.33(1), &
&
%decomposes as -0.34 = -0.03 - 0.17 - 0.13
\end{align}
so that, due to the resonance enhancement, IB effects as large as $30\%$ are observed, and the mixed $\Order(e^2\delta)$ contributions come out larger than in the $2\pi$ channel. 
While the 
$K^0$ self energy is negligible, indirect $\Order(e^2)$ effect from the $K^\pm$ contribution to the $\phi$ spectral function still produce a non-vanishing value in Eq.~\eqref{KKbar}, and the remaining differences between isospin-limit $K^+K^-$ ($16.29$) and $\bar K^0 K^0$ ($16.47$) are due to $c_\phi$ and the isovector form factor.

\section{Phenomenological estimates of isospin-breaking effects in the HVP contribution}

\nocite{James:2021sor}

A summary of all effects is shown in Table~\ref{tab:IB}, separately for $\Order(e^2,\delta)$ contributions and the decomposition into Euclidean windows. The comparison to Refs.~\cite{Borsanyi:2020mff,Blum:2018mom} indicates somewhat larger values in the intermediate window, especially for $\Order(e^2)$, but we emphasize that the quoted uncertainties in our phenomenological estimates do not include effects from the missing exclusive channels, which are expected to become most relevant in the intermediate and SD windows.\footnote{Estimating IB effects in subleading channels becomes increasingly challenging. In the $3\pi$ channel, threshold effects are strongly suppressed by phase space, while again IB in the residue $c_\omega^{3\pi}$ is hard to quantify. FSR effects should scale $\simeq 0.4$ by naively comparing to the $2\pi$ channel, and the dependence of $\Gamma_\omega$ on the pion mass~\cite{Dax:2018rvs} cancels out in the integral. Model-based estimates indicate $\simeq -0.6$ from a $\rho\to3\pi$ component~\cite{Boito:2022rkw,BABAR:2021cde}, but the underlying fit function cannot be reconciled with the analytic properties of the $\gamma^*\to 3\pi$ amplitude. QED corrections to the $R$-ratio are suppressed by $\Order(10^{-3})$~\cite{Harlander:2002ur}, which implies a correction $\lesssim 0.1$ in the HVP integral for the energy range in which perturbative QCD applies.} For the full HVP contribution, both our estimate and the inclusive ChPT determination from Ref.~\cite{James:2021sor}  (with the critical low-energy constant $\delta C_{93}^{(1)}$ extracted from $\tau$ decays) indicate a larger $\Order(\delta)$ effect, albeit largely consistent within uncertainties. For $\Order(e^2)$ we observe good agreement with Ref.~\cite{Borsanyi:2020mff}, which emerges as a result of substantial cancellations among several individually large effects.

\begin{table}
 \begin{center}
 \scalebox{0.788}{
\begin{tabular}{crrrrrrrr}
\toprule
& \multicolumn{2}{c}{SD window} & \multicolumn{2}{c}{int window} & \multicolumn{2}{c}{LD window} & \multicolumn{2}{c}{full HVP}\\
& $\Order(e^2)$ & $\Order(\delta)$& $\Order(e^2)$ & $\Order(\delta)$& $\Order(e^2)$ & $\Order(\delta)$& $\Order(e^2)$ & $\Order(\delta)$\\\midrule
$\pi^0\gamma$ & $0.16(0)$ & -- & $1.52(2)$  & -- & $2.70(4)$ & -- & $4.38(6)$ & --\\
$\eta\gamma$ & $0.05(0)$ & -- & $0.34(1)$ & -- & $0.31(1)$ &-- & $0.70(2)$ & --\\
$\rho$--$\omega$ mixing & -- & $0.05(0)$ & -- & $0.83(6)$ & -- & $2.79(11)$ & -- & $3.68(17)$\\
FSR ($2\pi$) & $0.11(0)$ & -- & $1.17(1)$ & -- & $3.14(3)$ &-- & $4.42(4)$ & --\\
$\mpii$ vs.\ $\mpic$ ($2\pi$) & $0.04(1)$ & -- & $-0.09(7)$ & -- & $-7.62(14)$ & --& $-7.67(22)$ &--\\
FSR ($K^+ K^-$)  & $0.07(0)$ & -- & $0.39(2)$ & -- & $0.29(2)$ &-- & $0.75(4)$ & --\\
kaon mass ($K^+ K^-$)  & $-0.29(1)$ & $0.44(2)$ & $-1.71(9)$ & $2.63(14)$ & $-1.24(6)$ & $1.91(10)$ & $-3.24(17)$ & $4.98(26)$\\
kaon mass ($\bar K^0 K^0$)  & $0.00(0)$ & $-0.41(2)$  & $-0.01(0)$ & $-2.44(12)$ & $-0.01(0)$ & $-1.78(9)$ & $-0.02(0)$ & $-4.62(23)$\\\midrule
total  & $0.14(1)$ & $0.08(3)$ & $1.61(12)$ & $1.02(20)$ & $-2.44(16)$ & $2.92(17)$ & $-0.68(29)$ & $4.04(39)$\\\midrule
Ref.~\cite{Borsanyi:2020mff} & -- & -- & $-0.09(6)$ & $0.52(4)$ & -- & -- & $-1.5(6)$ & $1.9(1.2)$\\
Ref.~\cite{Blum:2018mom} & -- & -- & $0.0(2)$ & $0.1(3)$ & -- & -- & $-1.0(6.6)$ & $10.6(8.0)$\\
Ref.~\cite{James:2021sor} & -- & -- & -- & -- & -- & -- & -- & $3.32(89)$\\
\bottomrule
\end{tabular}
}
\end{center}
\caption{Summary of IB effects from $\pi^0\gamma$, $\eta\gamma$, $2\pi(\gamma)$, and $\bar K K(\gamma)$, separated into short-distance (SD), intermediate, and LD window, in comparison to the lattice-QCD calculations from Refs.~\cite{Borsanyi:2020mff,Blum:2018mom} and the ChPT estimate of the $\Order(\delta)$ contribution from Ref.~\cite{James:2021sor}.}
\label{tab:IB}
\end{table}

\section{Conclusions}
\label{sec:conclusions}

In this contribution we collected phenomenological estimates of IB effects in the HVP contribution to the anomalous magnetic moment of the muon, improving especially the pion-mass correction in the $2\pi$ channel and adding an estimate of resonance-enhanced $\bar K K$ effects. In particular, we provided a breakdown into $\Order(e^2,\delta)$ components and Euclidean windows, see Table~\ref{tab:IB} for the main results. Given the limitations of the phenomenological approach to obtain inclusive numbers, there is reasonable agreement with current lattice-QCD calculations. In some cases, $\Order(e^2)$ for the intermediate window and $\Order(\delta)$ for the full HVP contribution, some differences are observed, but in both cases the result would increase further if the phenomenological estimates were adopted. We thus conclude that IB corrections are unlikely to play a relevant role in understanding the tension between $e^+e^-$ data and lattice QCD.

\section*{Acknowledgments}
We thank Alexander Keshavarzi for a cross check of the window decomposition of the $\eta\gamma$ channel, as well as Laurent Lellouch and Antonin Portelli for valuable discussions on the kaon mass in different isospin schemes. 
Support by the SNSF (Project Nos.\ 200020\_175791, PCEFP2\_181117, and PCEFP2\_194272) and the DFG, through the funds provided to the Sino--German Collaborative
Research Center TRR110 ``Symmetries and the Emergence of Structure in QCD''
(DFG Project-ID 196253076 -- TRR 110), is gratefully acknowledged.

\bibliographystyle{JHEP}
\bibliography{ref}

\end{document}